# Interprétation vague des contraintes structurelles pour la RI dans des corpus de documents XML

## Évaluation d'une méthode approchée de RI structurée

**Eugen Popovici – Gildas Ménier – Pierre-François Marteau**

*VALORIA*
*Université de Bretagne Sud*
*Rue Yves Mainguy, 56 000 Vannes*
*{eugen.popovici, gildas.menier, pierre-francois.marteau}@univ-ubs.fr*

RÉSUMÉ. *Nous proposons des algorithmes dédiés à l'indexation et à la recherche approximative d'information dans les bases de données hétérogènes semi-structurées XML. Le modèle d'indexation proposé est adapté à la recherche de contenu textuel dans les contextes XML définis par les structures d'arbres. Les mécanismes de recherche approchée mis en œuvre s'appuient sur une distance de Levenshtein modifiée et des heuristiques de fusion d'information. Une implémentation exploitant simultanément l'information structurée, i.e. l'arborescence des éléments XML, et le contenu des documents indexés est décrite. Les performances obtenues dans le cadre de la campagne d'évaluation INEX 2005 sont présentées et analysées. Celles-ci positionnent l'approche proposée parmi les meilleurs systèmes évalués, sur la tâche de recherche approximative de contenu en contexte structurel vague.*

ABSTRACT. *We propose specific data structures designed to the indexing and retrieval of information elements in heterogeneous XML data bases. The indexing scheme is well suited to the management of various contextual searches, expressed either at a structural level or at an information content level. The approximate search mechanisms are based on a modified Levenshtein editing distance and information fusion heuristics. The implementation described highlights the mixing of structured information presented as field/value instances and free text elements. The retrieval performances of the proposed approach are evaluated within the INEX 2005 evaluation campaign. The evaluation results rank the proposed approach among the best evaluated XML IR systems for the VVCAS task.*

MOTS-CLÉS : *XML, Base de données hétérogènes, recherche et extraction d'information, fusion d'information, distance d'édition de Levenshtein, opérateurs de recherche, INEX.*

## 1. Introduction

La société de l'information engendre un nombre considérable de documents hétérogènes, tant en forme qu'en nature. Face à cette hétérogénéité, l'émergence d'XML traduit un effort croissant qui vise à normaliser la présentation et l'archivage de l'information. Dans nombre de domaines de spécialité des structures de documents standardisés (*Document Type Definition*) et des langages d'interrogation *centrés données* (*XPath*, *XQuery*) sont proposées pour améliorer le partage, l'échange d'information et l'interopérabilité des applications. Cet effort de normalisation, s'il permet d'espérer une amélioration d'accès aux données disponibles en ligne ne résout pas cependant l'ensemble des difficultés concernant la qualité de l'information accessible. Pour répondre à ce besoin, les approches *centrées documents* visent le développement d'algorithmes de recherche d'information structurée basés sur une exploitation des contextes structurels d'occurrence des éléments informationnels et sur des heuristiques de classement des réponses. Parmi ces dernières on peux citer des approches basées sur : le modèle probabiliste (Fuhr *et al.*, 2004) (Lalmas *et al.*, 2000) ; le modèle de langage (Abolhassani *et al.*, 2004) (Kamps *et al.*, 2004) (Ogilvie *et al.*, 2006) ou sur l'exploitation des réseaux bayésiens (Piwowarski *et al.*, 2004). Les approches les plus répandues restent cependant les différentes adaptations et extensions du modèle vectoriel (Geva, 2006) (Grabs *et al.*, 2002) (Hubert, 2005) (Mass *et al.*, 2005) (Sauvagnat *et al.*, 2004) (Trotman, 2005).

*W3C* propose *XQuery 1.0 and XPath 2.0 Full-Text* (Amer-Yahia *et al.*, 2005) s'appuyant sur les approches « centrées données » et sur des techniques des classements du contenu textuel de documents. Cependant, les contraintes structurelles pour la portée de la recherche d'information textuelle doivent être complètement spécifiées et leur interprétation reste toujours stricte. Parmi le nombre d'approches développées pour la recherche d'information structurée il existe peu de modèles qui traitent de la structure des documents d'une manière approchée. La plupart des méthodes proposent simplement de filtrer (ou pondérer) les résultats textuels en fonction des contraintes structurelles (Theobald *et al.*, 2002). Du fait de l'hétérogénéité des structures et des contenus des documents disponibles au format XML, il n'est pas envisageable d'imaginer que l'utilisateur puisse disposer d'une connaissance suffisante sur la structure des documents accessibles pour orienter sa recherche d'information.

Différentes approches traitant d'une manière flexible les contraintes structurelles ont été proposées pour rentabiliser l'effort de structuration dans le cadre de la recherche d'information XML. Ces approches sont basées sur des relations sémantiques (Theobald *et al.*, 2002) ou sur différents degrés de relaxation concernant l'organisation des informations structurelles. Parmi ces dernières on peut citer des techniques basées sur des correspondances d'arbres (Schlieder, 2002) (Amer-Yahia *et al.*, 2004) ou des chemins (Carmel *et al.*, 2003) (Amer-Yahia *et al.*, 2005b).

Les algorithmes d'alignement approximatif d'arbres utilisant des distances d'éditions classiques (Tai, 1979) sont caractérisés par une haute complexité algorithmique. L'alignement d'arbres ordonnés a une complexité de $O(|N_1| \cdot |N_2| \cdot prf(T_1) \cdot prf(T_2))$, où $|N_i|$ est le nombre de nœuds contenus dans l'arbre $T_i$ et $prf(T_i)$ la profondeur de l'arbre $T_i$, - sachant que l'alignement d'arbres non ordonnés est un problème NP-difficile (Zhang *et al.*, 1992). Les algorithmes d'alignement décrits dans la littérature sont basés sur des distances d'édition et de fonction d'association ou de correspondance. La complexité de ces algorithmes d'alignement peut être améliorée en introduisant des contraintes sur l'ordre des séquence d'opérations d'édition, ou sur les opérations autorisées (Schlieder, 2002) (Amer-Yahia *et al.*, 2004). La complexité algorithmique de ces principes d'alignement est trop élevée au regard du grand volume documentaire qu'il est nécessaire de considérer pour une exploitation à grande échelle. L'expressivité et la complexité des requêtes exprimées sous forme d'arbres ou des chemins pour la recherche approchée d'informations structurées ont été étudiées dans (Amer-Yahia *et al.*, 2005b). Les auteurs montrent de façon expérimentale que les approches basées sur la correspondance des chemins fournissent des résultats ayant une qualité comparable en précision à celles basées sur la correspondance d'arbres tout en améliorant les performances du système.

Dans cet article, nous proposons plus précisément d'évaluer l'amélioration de l'accès à l'information apportée par des mécanismes d'alignement approximatif des chemins structurels (Menier *et al.*, 2002) (Marteau *et al.*, 2003) similaires à ceux introduits dans (Carmel *et al.*, 2003). Nous décrivons également un modèle de recherche d'information inspiré du modèle vectoriel (Salton *et al.*, 1988) susceptible de s'appuyer sur les éléments structurants des documents semi-structurés pour affiner l'analyse et la recherche des contenus non structurés. Les résultats concernant la pertinence de l'approche obtenus dans le cadre de INEX 2005 (*INitiative for the Evaluation of XML Retrieval*[1]) montrent l'intérêt de notre proposition.

## 2. Indexation en Contexte XML

Un document XML est représentable sous la forme d'une structure arborescente ordonnée dont les nœuds correspondent à des éléments XML. Les nœuds et en particulier les feuilles de l'arbre peuvent contenir des éléments d'information textuels ou binaires (TEXT ou CDATA) ou faire référence à des éléments externes tels que des images, des vidéos ou du texte. Un exemple de document XML extrait de la collection *inex-1.8* associé à son arbre est présenté en Figure 1.

---

[1] http://inex.is.informatik.uni-duisburg.de

## 2.1. Contexte XML

Dans la structure de l'arbre codant un document XML, chaque nœud $n$ peut être rattaché à la racine de l'arbre $n_0$ par l'intermédiaire du chemin $p(n)$. Ce chemin est une séquence ordonnée de nœuds d'éléments XML $n_i$ éventuellement associes à un ensemble non ordonné de couples <attribut, valeur> $A(n_i)$. Il constitue le contexte d'occurrence du nœud $n$ (le nœud extrémité du chemin $p(n)$ ) dans le document considéré. Dans le cas où un nœud[2] de l'arbre peut être décomposé en sous éléments textuels $\{M_i\}$ (mot, lemme ou chaîne de caractères), chaque sous élément $M_i$ est considéré comme un nœud terminal ne possédant pas d'attribut. Dans ce cas, on considèrera que le contexte d'occurrence de $M_i$ est le chemin suivant :

$$p(n) = <n_0, A(n_0)> <n_1, A(n_1)> \ldots <n, A(n_n)> <M_i, \emptyset> .$$

**Figure 1.** *Exemple de fichier XML associé à son arbre extrait du corpus inex-1.8.*

---

[2] Feuille avec du contenu textuel ou nœud avec du contenu mixte.

### 2.2. Modèle d'indexation

Le processus d'indexation est basé sur des listes inverses adaptées pour la prise en compte des contextes XML. Pour ce modèle, les entrées de la liste inverse sont représentées par les sous éléments $M_i$ des nœuds $n$ de l'arbre associé au document. Pour un sous élément $M_i$ d'un nœud $n$, quatre informations sont rattachées :

– un lien vers l'adresse permettant de localiser le document (*Uniform Resource Locator : URL*). Ce lien permet de retrouver la source du document indexé,

– un lien vers le contexte XML *p(n)* caractérisant l'occurrence du nœud $n$ dans l'arbre du document. Ce contexte est utilisé pour le traitement des contraintes structurelles.

– un index spécifiant la localisation de l'élément $M_i$ à l'intérieur du document. Cet index est utilisé pour effectuer des traitements portant sur le contenu textuel de documents comme la recherche de phrases ou la recherche des mots ayant des positions proches dans le texte.

– un intervalle $\{n^{ID}, Max(n^{ID}, \{n_i^{ID}\})\}$ ou :

– $n^{ID}$ représente l'identificateur de nœud $n$ obtenu à l'aide d'un parcours préfixé de l'arbre ; et

– $\{n_i^{ID}\}$ l'ensemble des identificateurs des nœuds descendants de $n$.

Cet intervalle sert à la résolution des relations d'héritage des nœuds dans la structure arborescente du document d'une manière similaire à celle introduite dans (Grust, 2002).

## 3. Paradigme de recherche d'information

Du fait de l'hétérogénéité des structures et des contenus des documents disponibles au format XML, il n'est pas envisageable d'imaginer que l'utilisateur puisse être à même de connaître l'ensemble des structures correspondantes. Par suite, l'utilisateur n'est pas à même de spécifier de manière systématique une requête précise, tant dans sa structure que dans son contenu sémantique. Dans une telle situation, il semble raisonnable de proposer à la fois des principes de recherche d'information exacts et approchés pour gérer au mieux l'incertitude inhérente à la requête de l'utilisateur. Le format des documents indexés étant XML, il n'est pas restrictif de concevoir que la requête elle même puisse être représentée par un arbre: la requête peut être d'une manière générale traduite sous la forme de document XML. La recherche de bas niveau peut alors être appréhendée par le biais d'algorithmes d'alignement approximatif d'arbres, dans le but de faire correspondre de manière approchée certaines branches de l'arbre issu de la requête avec les arbres issus des documents indexés.

Plutôt que l'exploitation de principes d'alignement approximatif d'arbres (Schlieder, 2002) (Amer-Yahia *et al.*, 2004) de haute complexité algorithmique on recherche plutôt ici des principes de recherche de bas niveau capables d'aligner approximativement des sous chemins de type *p(n)* contenus dans les arbres associés aux documents indexés avec un ou des sous chemin *p(r)* contenus dans la spécification d'une requête. Nous recherchons à définir un langage de requête plus évolué, qui intègre des expressions de recherche élémentaires *{p(n)}* et des expressions plus complexes, constituées à partir de l'assemblage d'expressions élémentaires : elles exploitent des opérateurs booléens classiques ou ensemblistes, des opérateurs flous ou encore des opérateurs basés sur des heuristiques dédiées.

### 3.1. Alignement approximatif des sous-structures *p(n)*

Soit *R* une requête élémentaire exprimée sous la forme d'un mot $M_i$ associée à un chemin $p^R$ pour lequel les couples <attributs, valeur> sont remplacés par des conditions ou contraintes *Cd* qui portent sur les valeurs d'attributs.

$$p^R(n) = <n_0, Cd(n_0)> <n_1, Cd(n_1)> ... <n, Cd(n_n)> <M_i, \varnothing>$$

Pour permettre une interprétation flexible des contraintes structurelles, nous proposons d'évaluer la pertinence des réponses pour une requête élémentaire *R* par la similarité entre le chemin $p^R$ représentant la requête et l'ensemble des chemins indexés associé aux réponses de la manière suivante :

$$\sigma(p^R, p_i^{R,D}) = 1 / (1 + \delta_L (p^R, \{p_i^{R,D}\}))$$

où $\delta_L$ s'apparente à une pseudo distance d'édition de type Levenshtein (Levenshtein, 1966) et $\{p_i^{R,D}\}$ représente l'ensemble des chemins issus de la racine de l'arbre $T^D$ associé au document *D* et dont le nœud extrémité correspond de manière stricte ou approximative au nœud extrémité du chemin $p^R$ correspondant à la requête.

Pour $p^R$ et $p_i^{R,D}$ donnés, la similarité $\sigma(p^R, p_i^{R,D})$ tend vers 0 quand la distance $\delta_L (p^R, p_i^{R,D})$ augmente. Naturellement, pour une concordance totale on aura une similarité maximale $\sigma(p^R, p_i^{R,D}) = 1$. La similarité structurelle basée sur la formule de normalisation introduite ci-dessus pénalise d'une manière non uniforme les écarts mis en évidence par la distance d'édition. Cette fonction a été définie de telle sorte que le score associé aux chemins ayant une correspondance parfaite ou presque parfaite avec le chemin spécifié dans la requête soit amplifié au détriment des correspondances plus éloignées.

La complexité d'un tel algorithme est : $O(l(p^R) \cdot prf(T^D) \cdot |\{p_i^{R,D}\}|)$, avec $|\{p_i^{R,D}\}|$ le cardinal de l'ensemble des chemins $\{p_i^{R,D}\}$, $l(p^R)$ la longueur du chemin $p^R$ et $prf(T^D)$ la profondeur de l'arbre $T^D$.

Nous conjecturons que cette approche est adéquate pour le traitement de la plupart des pages web ou des documents qui y sont référencés principalement parce que la profondeur des arbres engendrés par ces données est relativement faible. Des études statistiques sur les documents XML accessibles sur le WEB (Mignet *et al.*, 2002) montrent que le niveau moyen de profondeur pour les arbres est de 4, et 99% des documents analysés ont une profondeur inférieure à 8.

### 3.2. Distance d'édition entre deux chemins $p^R$ et $p_i^{R,D}$ : $\delta_L(p^R, p_i^{R,D})$

Soit $p^R$ le chemin correspondant à la requête élémentaire $R$ et $\{p_i^{R,D}\}$ l'ensemble des chemins retenus pour la recherche de similarité. Nous proposons une pseudo-distance[3] d'édition (Wagner *et al.*, 1974) en exploitant une matrice de coût paramétrable pour calculer la similarité entre un chemin $p_i^{R,D}$ et le chemin correspondant à la requête élémentaire $p^R$ (Ménier *et al.*, 2002). Cette distance détermine la transformation de coût minimal qui permet de transformer $p_i^{R,D}$ en $p^R$.

Les opérations élémentaires d'édition exploitées par cette distance et la fonction de coût $c$ qui leur est associée sont définies de la manière suivante :

- **La substitution** : un nœud $n^R$ du chemin $p^R$ est remplacé par (ou aligné sur) un nœud $n$ du chemin $p_i^{R,D}$ avec un coût élémentaire $c(n^R, n)$ que l'on considère pouvoir varier entre *0* (correspondance parfaite) et $c^{max} = 1$ (aucune correspondance).
- **L'élimination** : un nœud $n$ du chemin $p_i^{R,D}$ est éliminé avec un coût $c(n, \varepsilon)$ que l'on considère pouvoir varier entre *0* et $c^{max} = 1$ .
- **L'insertion** : un nœud $n$ est inséré dans le chemin $p_i^{R,D}$ avec un coût $c(\varepsilon, n)$ que l'on considère pouvoir varier entre *0* et $c^{max} = 1$ .

Des heuristiques complexes peuvent être exploitées pour définir le coût de substitution $c(n^R, n)$ en prenant en compte :

- des relations sémantiques au niveau des identificateurs d'éléments XML (par exemple 'chapitre', 'section') ;
- et/ou le degré de satisfaction approximative de conditions ou contraintes portant sur des attributs (Marteau *et al.*, 2003).

Soit l'ensemble $T_{p_i^{R,D} \to p^R}$ des transformations décomposables en une séquence de transformations élémentaires (insertion, élimination ou substitution) et permettant de transformer le chemin $p_i^{R,D}$ en $p^R$. A toute transformation $\tau$ de l'ensemble $T_{p_i^{R,D} \to p^R}$ on associe un coût global noté $C(\tau)$ assimilé à la somme des coûts des transformations élémentaires qui composent $\tau$. L'algorithme de Wagner&Fisher (Wagner *et al.*, 1974) permet de déterminer parmi les transformations de l'ensemble $T_{p_i^{R,D} \to p^R}$ la transformation de coût minimum notée $\tau^*$ (transformation optimale) avec une complexité $O(l(p_i^{R,D}) \cdot l(p^R))$ comme précisé précédemment.

Soit $C_i^*$ le coût minimum d'alignement du chemin $p_i^{R,D}$ sur le chemin $p^R$ :

---

[3] On parle de pseudo distance dans la mesure où la propriété de symétrie n'est pas assurée quelle que soit la fonction de coût $c$ envisagée.

$$\delta_L(p^R, p_i^{R.D}) = C_i^* = \underset{\tau \in Tp_i^{R.D} \to p^R}{Min} C(\tau) = C(\tau^*) \, .$$

Pour $p^R$ et $p_i^{R.D}$ donnés, la similarité $\sigma(p^R, p_i^{R.D}) \to 0$ quand le nombre des nœuds différents augmente. Naturellement, pour une concordance totale (mêmes éléments XML avec conditions satisfaites sur les attributs) on aura un coût minimum $C_i^* = 0$ et une similarité maximale $\sigma(p^R, p_i^{R.D}) = 1$.

### 3.3. Requêtes complexes

Des requêtes complexes sont construites à partir de requêtes élémentaires et d'opérateurs de composition booléens ou relevant d'heuristiques spécifiques : à ce jour, nous proposons huit opérateurs pour la construction des requêtes hors prise en compte des attributs. Les opérateurs pour la spécification des contraintes portant sur les attributs sont définis et détaillés dans (Marteau *et al.*, 2003). Les opérateurs utilisés pour la construction du langage de requête sont les suivants :

– les opérateurs booléens ou ensemblistes n-aires {*or, and*} qui permettent de rechercher une liste d'arguments dans un même contexte XML,

– l'opérateur ensembliste binaire {*without*} qui permet d'éliminer de la liste des résultats en cours, les résultats d'une autre requête,

– l'opérateur permettant de rechercher une séquence d'arguments {*seq*} : par exemple, (seq message * erreur) recherchera les éléments contenant les mots 'message' suivi par un mot quelconque (*), suivi par le mot 'erreur',

– l'opérateurs de recherche en contexte : {*in*} qui permet de rechercher une liste d'arguments respectivement dans un contexte XML déterminé,

– l'opérateur de recherche structurelle approchée {*in+*} permettant de fusionner par le biais d'une fonction d'agrégation linéaire pondérée les valeurs de similarité obtenues sur les informations structurelles et sur le contenu,

– l'opérateur {*same+*} permettant de pondérer les arguments de l'opérateur en fonction d'une heuristique prédéfinie, de type TFIDF (*Term Frequency by Inverse Document Frequency*) (Salton *et al.*, 1988) par exemple.

– l'opérateur ensembliste binaire {*filter*} permettant de filtrer un ensemble d'éléments résultats à partir d'un autre ensemble de résultats tenant compte de la pertinence associée à leur contexte global d'occurrence.

Le système analyse une requête *R* et produit un ensemble de réponses pondérées par un degré de pertinence. Soit $r(R) = \{ (e_i, v_i) \}$ l'ensemble réponse produit, avec $e_i$ un élément de réponse et $v_i$ une valeur numérique prise dans *[0..1]* indiquant sa pertinence par rapport à la requête.

Une requête complexe est en fait un arbre de requêtes, dans lequel l'ensemble final est fabriqué par fusion (ou filtrage) successive des ensembles créés par les requêtes filles $R_n$. Une requête fille peut être elle-même une requête complexe, ou bien une requête simple qui ne met en œuvre qu'un mot associé à un contexte d'occurrence recherché $p^R$ ou un mot seul ($M_i$). Ces combinaisons ensemblistes font intervenir des opérateurs de modification de pertinence. En nous inspirant des principes de décision multicritères floue (Zadeh, 1965) (Yager, 1977) nous avons défini les opérateurs suivants :

$$r( or\ (R_0,...R_n)\ ) = \{\ (e_i, v_i)\ \}\ \text{avec}\ v_i = \underset{k}{Max}(v_k)\ \text{tq}\ (e_i, v_k) \in \overset{n}{\underset{j}{\bigcup}} r(R_j)\ ;$$

$$r( and\ (R_0,...R_n)\ ) = \{\ (e_i, v_i)\ \}\ \text{avec}\ v_i = \underset{k}{Min}(v_k)\ \text{tq}\ (e_i, v_k) \in \overset{n}{\underset{j}{\bigcup}} r(R_j)\ ;$$

$$r( without\ (R_0,\ R_1)\ ) = \{\ (e_i, v_i)\ \}\ \text{avec}\ v_i = Max(0, v_i^0 - v_i^1)\ \text{tq}$$

$(e_i, v_i^0) \in r(R_0)$ et $(e_i, v_i^1) \in r(R_1)$ ;

$r( seq\ (M_0,...M_n)\ ) = \{\ (e_i, v_i)\ \}$, $v_i = 1$ si la séquence $M_0..M_n$ existe dans l'élément $e_i$, $0$ si non ;

$$r( in\ (path\ ,\ R_0,...R_n)) = \{\ (e_i, v_i)\ \}\ \text{avec}\ v_i = Min\{\ \underset{k}{Min}\ (v_k),\ \Delta(path, p(e_i))\}\ \text{tq}$$

$(e_i, v_k) \in \overset{n}{\underset{j}{\bigcup}} r(R_j)$ ; ou $p(e)$ le chemin de l'élément $e$ ; et $\Delta\ (path, p(e))=1$ ssi

$path==p(e)$, $0$ si non ;

$$( in+ (path,\ R_0,...R_n)) = \{\ (e_i, v_i)\ \}\ \text{avec}\ v_i = \beta.\Delta(\ path, p(e_i)) + (1 - \beta).\underset{k}{Min}(v_k)$$

tq $(e_i, v_k) \in \overset{n}{\underset{j}{\bigcup}} r(R_j)$, ou $\Delta\ (path, p(e)) = \sigma\ (path, p(e))$ la similarité structurelle entre les deux chemins (section 3.1) et $\beta \in [0..1]$ un facteur permettant de pondérer l'importance des informations sur la structure des documents vis-à-vis de leur contenu ;

$$r( same+ (R_0,...R_n)) = \{\ (e_i, v_i)\ \}\ \text{avec}\ v_i = \tau \cdot \underset{k}{\sum} \lambda_k.v_k\ \text{tq}\ (e_i, v_k) \in \overset{n}{\underset{j}{\bigcup}} r(R_j)\ ;$$

ou $\lambda_k$ est un facteur de pondération qui caractérise le pouvoir discriminant de l'élément de réponse $(e_i, v_k)$ dans la collection :

$\lambda_k = 1 - log(\ (1 + N^{D(e_i, v_k)})\ /\ (1 + N^D)\ )$ ; ou $N^{D(e_i, v_k)}$ le nombre de documents contenant l'élément de réponse $(e_i, v_k)$ ; $N^D$ le nombre total de documents ; et $\tau$ une constante de normalisation $\tau = 1\ /\ \Sigma_k(\lambda_k)$ ;

Soit $e^D$ un élément de réponse descendant du document $D$.

$$r( filter\ (R_0,\ R_1)) = \{\ (e_i^D,\ v_i)\ \}\ \text{avec}\ v_i = (v_i^1 + \underset{j}{Max}(v_j^0))/2\ \text{tq}$$

$\forall_D \forall_i (e_i^D, v_i^1) \in r(R_1)$ et $\forall_j (e_j^D, v_j^0) \in r(R_0)$ .

## 4. Évaluations

L'approche décrite ci-dessus a été évaluée dans le cadre de la tache *ad-hoc* lors de la campagne INEX 2005 (Popovici *et al.*, 2006). Afin d'évaluer la performance des différents SRI pour la RI structurée, INEX met à disposition des participants une collection de tests, de requêtes portant sur la structure et le contenu de documents ainsi que les jugements de pertinence associés.

Dans la suite, nous présentons des évaluations concernant l'évolution du temps d'indexation en fonction du volume de données traité, ainsi que le temps de réponse et les performances concernant la qualité des accès aux informations.

### 4.1. Les données expérimentales

La collection de test *inex-1.8* contient 16819 articles en format XML provenant de 24 revues de *IEEE Computer Society* publiées entre 1995-2004. Cette base totalise un volume de 725 Mo dans sa forme canonique. La collection est considérée sans attributs ou avec des valeurs d'attributs sans intérêt pratique pour les utilisateurs (Trotman *et al.*, 2004 – « Une note sur les attributs »). Elle contient 141 noms d'éléments XML différents composant 7948 contextes XML uniques en ignorant les attributs et les valeurs des attributs. La longueur maximale d'un contexte est de 20 nœuds, avec une longueur moyenne de 8 nœuds[4].

### 4.2. Le jeu de requêtes

Dans le cadre d'INEX sont introduites deux types de requêtes : les requêtes de type CO (*Content Only*) qui décrivent uniquement le contenu souhaité des éléments XML recherchés ; et les requêtes de type CAS (*Content And Structure*) qui portent sur la structure et le contenu de documents.

Le jeu de requêtes *INEX 2005 CAS* est formé de 47 requêtes contenant des références explicites sur la structure de documents. Par exemple, la requête « 280 » (Figure 2) recherche des *sections* décrivant des algorithmes approchés dans des *articles* contenant une référence bibliographique *bb* à Baeza-Yates ou des *sections* sur la correspondance des chaînes de caractères.

---

[4] Les statistiques sont calculées du point de vue du système d'indexation. C'est-à-dire, en utilisant des classes de noms d'éléments XML équivalentes pour les *Paragraphs*, *Sections*, *Lists*, et *Headings* comme défini dans (Trotman *et al.*, 2004). De la même façon, les contextes XML associés aux éléments vides ne comptent pas dans les résultats.

```
<inex_topic topic_id="280" query_type="CAS" ct_no="137">
    <InitialTopicStatement>find sections describing ways to use approximate string matching
    </InitialTopicStatement>
    <title/>
    <castitle>//article[ about(.//bb, Baeza-Yates) and about(.//sec , string matching)]
                //sec[about(., approximate algorithm)]</castitle>
    <description>find sections about approximate algorithms in works about string matching
        citing Baeza-Yates.</description>
    <narrative>...</narrative>
</inex_topic>
```

**Figure 2.** *Un exemple de requête complexe sur la structure et le contenu - topic 280 du jeu de requêtes INEX 2005 CAS exprimée en NEXI (Trotman et al., 2004).*

Parmi ces requêtes, 30 sont des requêtes subordonnées de la forme //A[B][5], 6 sont des requêtes indépendantes de la forme //A[B] et 11 sont des requêtes complexes de la forme //A[B]//C[D][6] constituées des requêtes subordonnées. Seulement 17 requêtes CAS – les requêtes complexes et les requêtes indépendantes – sont utilisées pour l'évaluation dans la campagne INEX 2005.

### 4.3. Mesures de pertinence

Les jugements de pertinence pour chaque requête sont effectués par les différents participants selon deux dimensions : *l'exhaustivité* et *la spécificité*. Un élément est exhaustif s'il contient toutes les informations requises par la requête ; un élément est spécifique si tout son contenu concerne la requête. Ces deux dimensions sont agrégées en utilisant deux types de fonction d'agrégation : *aggrégation stricte* (uniquement les réponses « parfaites » - complètement exhaustives et complètement spécifiques sont prises en compte) et *aggrégation généralisée* (qui récompense aussi des réponses partiellement pertinentes).

Les mesures officielles utilisées dans INEX 2005 sont : le gain cumulé étendu normalisé *(nxCG)* et la courbe effort-précision/gain-rappel *(ep/gr)*.

Pour un rang *i* donné, la valeur *nxCG[i]* reflète le gain relatif que l'utilisateur a accumulé jusqu'à ce rang par rapport au gain qu'il aurait pu obtenir si le système avait produit un classement optimal de réponses.

L'effort-précision (*ep*) à un niveau donné de gain-rappel (*gr*) indique l'effort nécessaire (estimé en nombre de réponses visitées) à un utilisateur parcourant les réponses retournées par un système, relatif à l'effort nécessaire pour accumuler le même niveau de gain-rappel utilisant un classement de réponses idéal. *MAep (Mean Average Effort Precision)* représente la moyenne non interpolée des valeurs

---

[5] //A[B] = retourne des éléments A à propos de B (Trotman *et al.*, 2004).
[6] //A[B]//C[D] = retourne des éléments C descendants de A, où A est à propos de B et C est à propos de D (Trotman *et al.*, 2004).

obtenues pour *ep* à des niveaux « naturels » de *gr* (i.e. à chaque réponse pertinente retournée).

Les détails sur les métriques utilisées pour l'évaluations dans le cadre de INEX 2005 sont disponibles dans (Kazai *et al.*, 2006).

### 4.4. Le processus d'indexation

La base de documents XML est transformée dans sa forme canonique[7]. Les mots fréquents et jugés non pertinents (les mots vides) sont éliminés. Au moment de l'indexation les mots sont *stemmés* (i.e. réduits à leur pseudo-radicaux) en utilisant l'algorithme de Porter (Porter, 1980). On indexe seulement les termes *alphanumériques* comme défini en (Trotman *et al.*, 2004). Les nombres, les attributs, les valeurs des attributs et les éléments XML vides ne sont pas indexés. Le modèle d'indexation est basé sur des listes inverses adaptées pour la prise en compte des contextes XML *p(n)* et sur un schéma d'étiquetage des nœuds permettant de résoudre les relations d'héritage dans l'arbre (Section 2.2) . Ce modèle est implémenté utilisant des structures de données de type BTree de Berkeley DB[8]. La taille de l'index obtenu représente 1,28 fois le volume des données initiales.

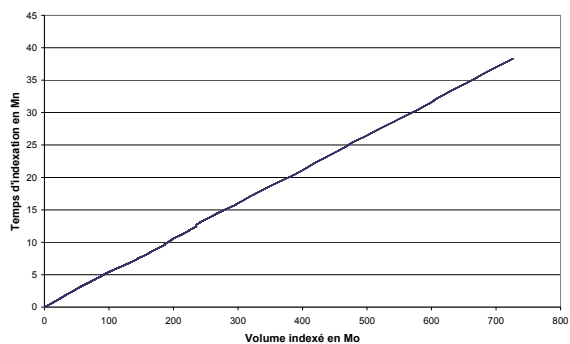

**Figure 3.** *Temps d'indexation en fonction de la taille des données indexées.*

La Figure 3 présente l'évolution du temps d'indexation (sur un Pentium IV - 2,4 GHz, avec 1,5 Go de mémoire RAM) pour la base inex-1.8 en fonction du volume des données indexées. Le temps nécessaire pour la création des listes inverses pour un volume de 725 Mo de données est d'environ 38 minutes. Cette évolution quasi linéaire montre que l'indexation de grandes bases documentaires XML est un objectif envisageable dans l'hypothèse des conditions indiquées. Néanmoins, on doit souligner que la prise en compte des attributs et de leurs valeurs multiplie le nombre de contextes possibles (7948 actuellement), ce qui complexifie la gestion de l'indexation des contextes entraînant une durée d'indexation nettement supérieure.

---

[7] Canonical XML Processor, http://www.elcel.com/products/xmlcanon.html
[8] http://www.sleepycat.com

### 4.5. Pondérations du schéma de similarité structurelle

Dans le langage officiel d'INEX 2005 – NEXI (Trotman *et al.*, 2004), le chemin associé a une requête est défini comme une séquence de nœuds parmi lesquels la seule relation acceptée est la relation de descendance. Par conséquence le chemin $p^R$ est modélisé comme une *sous séquence* de contextes indexés $p_i^{R,D}$, où une sous séquence ne doit pas nécessairement être constituée de nœuds contigus. Pour s'adapter à cette approche, on permet la suppression des nœuds dans les contextes indexés sans pénalisation en affectant le coût d'élimination d'un nœud dans la distance d'édition à $0$ : $c(n,\varepsilon) = 0$, $c(\varepsilon,n)=\xi$, $c(n^R,n)= \xi$ si $n^R{\neq}n$ où $0$ si $n^R{=}n$.

Pour illustrer le fonctionnement du mécanisme d'alignement on présente dans la Figure 4 les scores obtenus pour la distance d'édition $\delta_L$ $(p^R, \{ p_i^{R,D}\})$ entre le chemin correspondant à la requête 277 du jeu de requêtes INEX 2005 *//article[about(.//bb, Baeza-Yates)* et trois contextes XML extraits à partir de la collection *inex-1.8* (un extrait représentant la structure des documents dans la collection est donnée dans la Figure 1).

```
δ_L (//article//bb,   /article/bm/bib/bib1/bb/au/snm     ) =  0
δ_L (//article//bb,   /article/bm/app/bib/bib1/bb/au/snm ) =  0
δ_L (//article//bb,   /article/fm/au/snm                 ) =  1
```

**Figure 4.** *Exemples des distances entre le chemin « //article//bb » correspondant à la requête 277 du jeu de requêtes INEX 2005 et des contextes XML extraits de la collection inex-1.8.*

La requête recherche des *articles* ayant des références bibliographiques *bb* à *Baeza-Yates*. Dans les deux premiers cas, le chemin recherché *//article//bb* est une sous séquence de contextes indexés. Par conséquent la distance d'édition reste toujours égale à 0 indépendamment du fait que les deux contextes sont composés d'un nombre différent de nœuds. Dans le dernier cas, où *Baeza-Yates* est l'auteur de l'article, la distance d'édition va prendre une valeur égale à 1 pour pénaliser le fait que le nœud *bb* spécifié dans la requête n'apparaisse pas dans le contexte indexé.

La similarité structurelle basée sur le schéma de pondération ci-dessus modélise un utilisateur ayant une connaissance précise mais incomplète sur le nom des éléments XML indexés et sur leurs relations de descendance. Elle prend en compte : i) l'ordre d'occurrence des éléments correspondant entre la requête $p^R$ et le contexte indexé $p_i^{R,D}$ ii) et le nombre des éléments XML sans correspondant spécifié dans la requête $p^R$.

### 4.6. Processus de traitement de requêtes

#### 4.6.1  Traitement de requêtes orientées contenu (CO)

Les requêtes CO concernent les éléments qui ne contiennent que des éléments de recherche textuels. Nous calculons le score de pertinence pour tous les éléments feuille de l'arbre XML (avec au moins un des termes de recherché) en utilisant une variante du TF-IDF (voir la formule de pondération pour l'opérateur *same+*, Section 3.3). Dans notre approche, nous considérons comme valide, tout élément XML qui contient un terme recherché et ceci, indépendamment de sa taille.

#### 4.6.2  Traitement de requêtes orientées contenu et structure (CAS)

Les requêtes CAS regroupent deux cas :

– les requêtes simples de la forme *//A[B]* – c'est-à-dire la requête spécifie seulement les éléments cibles, et

– les requêtes complexes de la forme *//A[B]//C[D]* – la requête spécifie à la fois la cible (*//C[D]*) et le support (*//A[B]*).

##### 4.6.2.1   Traitement des éléments support et cible

Pour des requêtes simples de la forme //A[B] nous évaluons le contenu textuel des noeuds en utilisant le même principe que pour les requêtes CO. Les contraintes structurelles sont exploitées et interprétées comme des indices structurels (Trotman *et al.*, 2004). Nous calculons la similarité entre les contraintes structurelles de la requête et le chemin XML du fragment candidat en exploitant une distance d'édition modifiée (voir Section 4.5) qui fait intervenir des heuristiques dédiées aux attributs/valeurs (Marteau *et al.*, 2003). Enfin, nous fusionnons les résultats pour le contenu et la structure en utilisant une méthode de combinaison pondérée (voir la formule pour l'opérateur *in+*, Section 3.3).

##### 4.6.2.2   Traitement des conditions d'imbrication

En ce qui concerne les requêtes complexes du type *//A[B]//C[D]* (voir la Figure 2) nous évaluons la pertinence à la fois du support *//A[B]* et de la cible *//C[D]*. Nous ne conservons que les éléments cible qui possèdent au moins un élément support pertinent dans le même document. En effet, si un élément pertinent existe dans un document, son poids peut être propagé (en utilisant une fonction *Max*) vers la racine de l'arbre XML. Cet élément est un ancêtre (càd support) pour tous les éléments de l'arbre.

La mesure de similarité pour une requête complexe, met en oeuvre des modifications de pertinence associée à un élément résultat. Cette pertinence est calculée comme une moyenne arithmétique entre la pertinence de la cible et la pertinence maximum des supports (voir la formule de l'opérateur *filter*, Section 3.3). Ceci permet d'intégrer le concept de pertinence contextuelle du document dans le calcul de pertinence des éléments de réponses : les éléments cibles sans support sont écartés du résultat alors que ceux qui bénéficient d'un support très pertinent remontent dans le classement final.

Le classement définitif est trié par pertinence et les 1500 premiers résultats sont conservés.

### 4.7. Stratégies de recherche

Nous évaluons deux stratégies pour la recherche du contenu textuel : COCAS_SAMEPLUS (utilisant l'opérateur *same+*) et COCAS_SEQ (qui emploie une recherche séquentielle stricte (*seq*) à l'intérieur de l'opérateur *same+*).

Une recherche séquentielle stricte (*seq*) va filtrer les éléments ne contenant pas tous les mots recherchés en séquence avec des indices consécutifs dans le texte (ignorant les termes fréquents. Une recherche basée uniquement sur l'opérateur *same+* va retourner aussi des éléments contenant seulement un sous-ensemble des termes recherchés. Les résultats sont classés en prenant en compte :

- le nombre de termes différents recherchés trouvés à l'intérieur de l'élément,
- le caractère discriminant (i.e. IDF) des termes recherchés dans la collection.

Cette stratégie n'impose pas de contraintes sur l'ordre d'occurrence de termes recherchés.

Nous évaluons également l'effet produit par la prise en compte des informations structurelles dans le processus de recherche d'information sur la qualité de réponses retournées.

Une requête *CAS (Content And Structure)* exprime deux types de contraintes structurelles : celles qui spécifient les éléments XML dans lesquels il s'agit de rechercher l'information (les éléments supports) et celles qui spécifient les éléments XML qu'il faut retourner (les éléments cibles). Dans le cadre de la campagne INEX 2005 plusieurs tâches ont été proposées pour évaluer différents degrés d'approximation utilisés pour l'interprétation des contraintes structurelles portant sur les éléments cibles et support : VV (*Vague Target, Vague Support*), VS (*Vague Target, Strict Support*, SV (*Strict Target, Vague Support*) et SS (*Strict Target, Strict Support*).

Les stratégies VV, VS, SV et SS prennent en compte les informations explicites sur les contextes XML par le biais de l'opérateur *in+* avec $\beta=0,5$ pour une interprétation vague, et par le biais de l'opérateur *in* pour une interprétation stricte

(voir la Section 3.3). Pour la stratégie COCAS ces informations sont complètement ignorées. On peut remarquer que l'organisation hiérarchique des informations recherchées (càd. le rôle des éléments supports et cibles) est traitée d'une manière uniforme pour toutes les stratégies par l'utilisation de l'opérateur *filter*.

```
COCAS_SEQ        ( FILTER ( AND ( SAME+ ( SEQ  Baeza Yates ) ) )
                                 ( SAME+  string matching ) ) )
                 ( SAME+  approximate algorithm )
COCAS_SAMEPLUS   ( FILTER ( AND ( SAME+ Baeza Yates)
                                 ( SAME+ string matching ) )
                 ( SAME+ approximate algorithm ) )
VVCAS_SAMEPLUS   ( FILTER ( AND ( IN+ [/article/bb/]  ( SAME+ Baeza Yates ) )
                                 ( IN+ [/article/sec/] ( SAME+ string matching ) ) )
                 ( IN+ [/article/sec/] ( SAME+  approximate algorithm ) ) )
VSCAS_SAMEPLUS   ( FILTER ( AND ( IN  [/article/bb/]  ( SAME+ Baeza Yates ) )
                                 ( IN  [/article/sec/] ( SAME+ string matching ) ) )
                 ( IN+ [/article/sec/] ( SAME+  approximate algorithm ) ) )
SVCAS_SAMEPLUS   ( FILTER ( AND ( IN+ [/article/bb/]  ( SAME+ Baeza Yates ) )
                                 ( IN+ [/article/sec/] ( SAME+ string matching ) ) )
                 ( IN [/article/sec/] ( SAME+  approximate algorithm ) ) )
SSCAS_SAMEPLUS   ( FILTER ( AND ( IN [/article/bb/]  ( SAME+ Baeza Yates ) )
                                 ( IN [/article/sec/] ( SAME+ string matching ) ) )
                 ( IN [/article/sec/] ( SAME+  approximate algorithm ) ) )
```

**Tableau 1**. *Traduction de la requête complexe 280 utilisant des opérateurs du langage récursif (section 3.3) selon la stratégie de recherche adoptée.*

Chaque stratégie est basée sur un jeu de requêtes différent obtenu d'une manière automatique à partir du jeu de requêtes *INEX 2005 CAS* exprimé dans le langage NEXI (Trotman *et al.*, 2004). Selon les différentes interprétations, les transformations utilisent un ensemble spécifique d'opérateurs pour la description des requêtes. Un exemple de traduction pour la requête complexe 280 (Figure 2) selon les différentes stratégies de recherche est présente dans le Tableau 1.

### 4.8. Évaluation du temps de réponse pour la recherche d'information

La Figure 5 présente le temps de réponse du système (sur un Pentium IV - 2,4 GHz, avec 1,5 Go de mémoire RAM) estimé sur la base du jeu de requêtes complexes *INEX 2005 CAS* soumis sur la collection *inex-1.8* de 725 Mo.

Le temps moyen de réponse du système se situe entre 0,82 s pour une recherche séquentielle stricte qui ne nécessite pas de calcul pour la prise en compte des information structurelles, et 3,29 s pour une recherche séquentielle approximative basée sur l'opérateur *same+* et associée aux calculs sur la structure des documents.

Le coût en temps de réponse associé à la complexité des algorithmes de recherche approchée sur la structure des documents est non négligeable, mais pas prohibitif - particulièrement dans une perspective de parallélisation des algorithmes de recherche en vue de les exploiter dans un contexte de gestion de grandes masses de données.

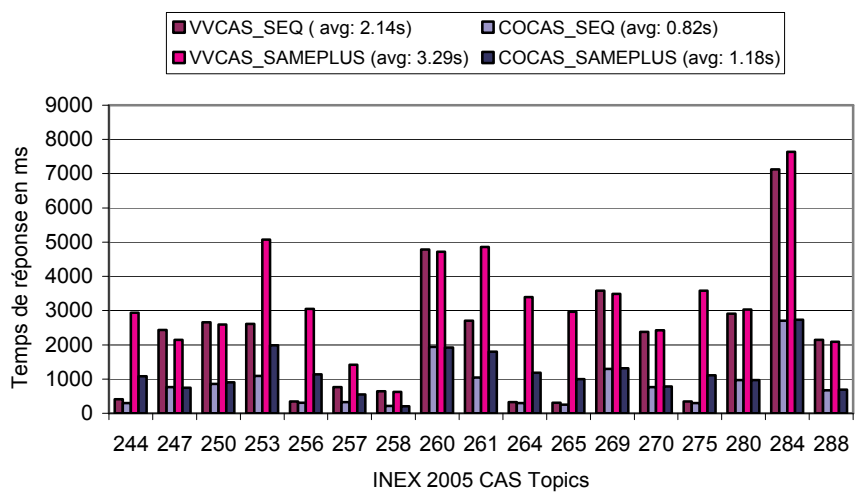

**Figure 5.** *Le temps de réponse pour 17 requêtes complexes extraites du jeu INEX 2005 CAS pour les différentes stratégies de recherches.*

### 4.9. Évaluation de la pertinence des réponses retournées

Dans la suite, nous présentons des résultats portant sur la qualité des réponses retournées par le système selon les différentes stratégies de recherche décrites, comparés aux résultats officiels d'INEX 2005 pour la tâche *VVCAS (Vague Target Vague Support Content And Structure)*. Pour la tâche *VVCAS* les contraintes structurelles sur les éléments supports et les éléments cibles sont interprétées comme des contraintes vagues.

Les métriques rapportées sont les mesures officielles utilisées dans INEX 2005 : la courbe effort-précision/gain-rappel *(ep/gr) et* le gain cumulé étendu normalisé *(nxCG)* calculés avec les paramètres *ov.=off*[9] *et quant.=strict*[10].

---

[9] les réponses entrelacées sont autorisées.
[10] seulement les éléments complètement exhaustives et spécifiques sont prises en compte.

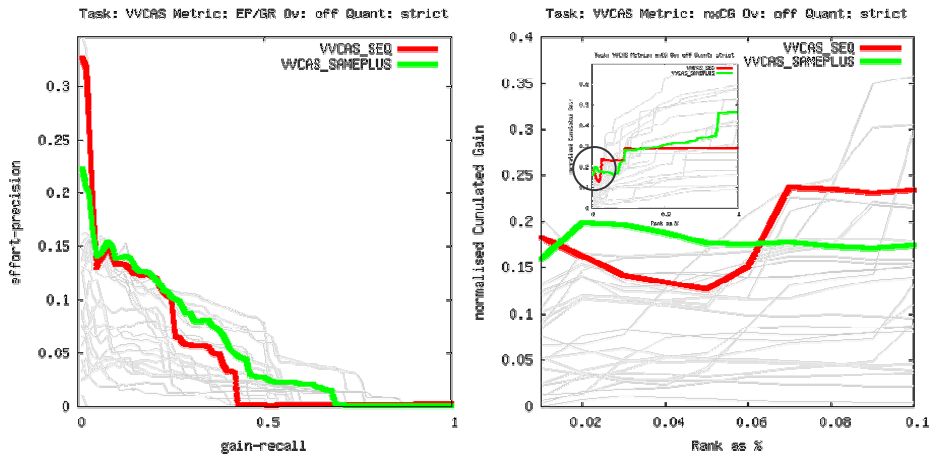

**Figure 6.** *Les courbes de rappel-précision ep/gr (gauche) et de gain cumulé étendu normalisé nxCG – zoom sur les premiers 150 réponses retournés - (droite) en autorisant les réponses imbriquées (Ov.=off) et utilisant une fonction d'agrégation stricte (Quant.=strict) pour les différentes stratégies de recherche comparées avec les résultats officiels d'INEX 2005, tâche VVCAS.*

La courbe *ep/gr* fournit une vue globale sur la performance des SRI et par conséquent est une mesure d'évaluation orientée système. *nxCG* est une mesure orientée utilisateur fournie pour évaluer les premiers résultats retournés par les SRI. Dans la Figure 6.b nous présentons une vue détaillée pour les premiers 150 résultats retournés.

En analysant les résultats obtenus (Tableau 2) comparés avec les résultats officiels de INEX 2005 (Figure 6) on en déduit : i) que le système a une bonne précision pour des valeurs de rappel peu élevées. En particulier, les meilleures valeurs rapportées pour *nxCG@{10,25,50} (overlap=off, quant.=strict)* (Tableau 2) peuvent nous classer (non officiellement) parmi les deux meilleurs (sur 28) soumissions pour la tâche VVCAS (Figure 6.b). ; ii) La meilleure performance globale est obtenue par la stratégie *VVCAS_SAMEPLUS* sur la mesure *ep/gr (ov.=off, quant.=strict)* avec *MAep = 0,0558*. Cette performance est équivalente à une cinquième place non officielle (Figure 6.a).

Les résultats rapportés dans le Tableau 2 montrent une augmentation de la qualité globale des résultats pour l'utilisation de la stratégie flexible pour la recherche de phrases (le gain sur *MAep* est de 14,58% pour *VVCAS_SAMEPLUS*). Cependant, cette stratégie conduit à une perte de précision (de 13,38% pour *nxCG@10* et de 31,65% pour *ep/gr à 0,01*) pour les premiers résultats retournés en comparaison avec l'approche stricte.

| | nxCG | | | ep/gr | | | | |
|---|---|---|---|---|---|---|---|---|
| *RunId* | *@10* | *@25* | *@50* | *0,01/@15* | *0,02/@30* | *0,03/@45* | *0,1/@150* | *MAep* |
| VVCAS_SAMEPLUS | 0,1444 | **0,2022** | **0,1889** | 0,2235 | 0,206 | 0,2013 | **0,1389** | **0,0558** |
| VVCAS_SEQ | **0,1667** | 0,1689 | 0,1339 | **0,327** | **0,3169** | **0,2673** | 0,133 | 0,0487 |
| Gain en % | -13,38 | 19,72 | 41,08 | -31,65 | -35 | -24,69 | 4,44 | 14,58 |

**Tableau 2.** *Le gain cumulé étendu normalisé nxCG[i] pour un rang i donné et l'effort-précision (ep) à un niveau donné de gain-rappel (gr) pour les différentes stratégies de recherche sur la tâche VVCAS en autorisant les réponses imbriquées et utilisant une fonction d'agrégation stricte. Les meilleurs résultats sont présentés en gras.*

Dans la Figure 7 et le Tableau 3 nous présentons des résultats portants sur l'interprétation vague ou stricte des contraintes structurelles sur les éléments cible et support (VV, VS, SV, SS). Les résultats sont comparés (Tableau 4) avec une stratégie orientée contenu CO qui ignore les informations explicitées sur la structure des documents mais prends en compte l'organisation hiérarchique des informations recherchées par le biais de l'opérateur *filter*[11]. Les différentes stratégies sont évaluées sur la tâche VVCAS (càd. en utilisant des jugements de pertinence non filtrés, dans la forme dans lesquelles ont été fournies par les utilisateurs).

| | nxCG | | | ep/gr | | | | |
|---|---|---|---|---|---|---|---|---|
| *RunId* | *@10* | *@25* | *@50* | *0,01/@15* | *0,02/@30* | *0,03/@45* | *0,1/@150* | *MAep* |
| SSCAS_SEQ | 0,0556 | 0,0711 | 0,0644 | 0,0476 | 0,0592 | 0,0646 | 0,0599 | 0,0217 |
| SVCAS_SEQ | 0,0556 | 0,0711 | 0,0644 | 0,0476 | 0,0592 | 0,0646 | 0,0599 | 0,0217 |
| VSCAS_SEQ | 0,1111 | 0,1022 | 0,0939 | 0,2702 | 0,2601 | 0,2036 | 0,0876 | 0,0358 |
| COCAS_SEQ | 0,1444 | 0,16 | 0,1273 | 0,3161 | 0,3084 | 0,2578 | 0,124 | 0,0448 |
| VVCAS_SEQ | **0,1667** | **0,1689** | 0,1339 | **0,327** | **0,3169** | **0,2673** | 0,133 | **0,0487** |
| SSCAS_SAMEPLUS | 0,0556 | 0,0711 | 0,0644 | 0,0476 | 0,0592 | 0,0646 | 0,0599 | 0,0259 |
| SVCAS_SAMEPLUS | 0,0556 | 0,0711 | 0,0644 | 0,0476 | 0,0592 | 0,0646 | 0,0599 | 0,0259 |
| VSCAS_SAMEPLUS | 0,0889 | 0,0933 | 0,0889 | 0,1595 | 0,1494 | 0,1397 | 0,0879 | 0,0372 |
| COCAS_SAMEPLUS | 0,1222 | 0,1933 | 0,1822 | 0,2126 | 0,1945 | 0,1936 | 0,1338 | 0,0525 |
| VVCAS_SAMEPLUS | **0,1444** | **0,2022** | **0,1889** | **0,2235** | **0,206** | **0,2013** | **0,1389** | **0,0558** |

**Tableau 3.** *nxCG[i] et ep/gr pour les différentes dégrées d'approximation utilisés pour l'interprétation des contraintes structurelles évaluées sur la tâche VVCAS. Les réponses imbriquées sont autorisées. Fonction d'agrégation stricte. Les meilleurs résultats sont présentés en gras.*

---

[11] Des comparaisons avec une stratégie CO qui prend en compte uniquement les informations sur le contenu des éléments sont effectuées sur la tâche COS.Focused d'INEX 2005 et présentées en Annexe.

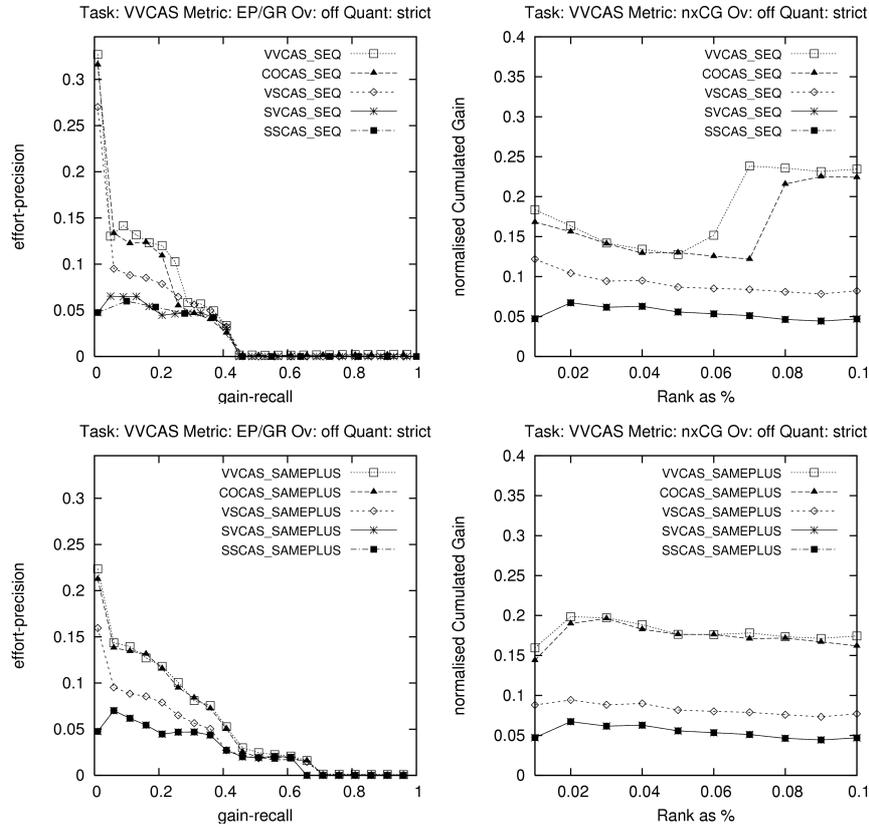

**Figure 7.** *Les courbes de rappel-précision ep/gr (gauche) et de gain cumulé étendu normalisé nxCG - pour les premiers 150 réponses retournés – (droite) pour les différents dégrés d'approximation utilisés pour l'interprétation des contraintes structurelles (VV, VS, SV, SS, CO) évaluées sur la tâche VVCAS d'INEX 2005. Les réponses imbriquées sont autorisées. Fonction d'agrégation stricte.*

En analysant les résultats présentés dans le Tableau 3 nous observons que les stratégies utilisant une interprétation stricte des éléments cibles - SV et SS - ont un comportement identique. Cette observation est confirmée par les expérimentations effectuées en (Trotman *et al.*, 2006a) portant sur la totalité des soumissions de la campagne INEX 2005. Par conséquent, pour l'analyse des résultats nous ferons référence uniquement à la stratégie SV.

| | | nxCG | | | ep/gr | | | | |
|---|---|---|---|---|---|---|---|---|---|
| | *RunId* | *@10* | *@25* | *@50* | *0,01/@15* | *0,02/@30* | *0,03/@45* | *0,1/@150* | *MAep* |
| SV, SS vs. CO | SAMEPLUS | -54,5 | -63,22 | -64,65 | -77,61 | -69,56 | -66,63 | -55,23 | -50,67 |
| | SEQ | -61,5 | -55,56 | -49,41 | -84,94 | -80,8 | -74,94 | -51,69 | -51,56 |
| VS vs. CO | SAMEPLUS | -27,25 | -51,73 | -51,21 | -24,98 | -23,19 | -27,84 | -34,3 | -29,14 |
| | SEQ | -23,06 | -36,13 | -26,24 | -14,52 | -15,66 | -21,02 | -29,35 | -20,09 |
| VV vs. CO | SAMEPLUS | 18,17 | 4,6 | 3,68 | 5,13 | 5,91 | 3,98 | 3,81 | 6,29 |
| | SEQ | 15,44 | 5,56 | 5,18 | 3,45 | 2,76 | 3,69 | 7,26 | 8,71 |

**Tableau 4.** *Gain en pourcentage introduit par les différents degrés d'approximation utilisés pour l'interprétation des contraintes structurelles (VV, VS, SV, SS) par rapport à la stratégie orientée contenu (CO) sur la tâche VVCAS.*

L'interprétation stricte des contraintes structurelles a un effet négatif sur la qualité des résultats de recherche - rejet des réponses globalement pertinentes mais partiellement/non pertinentes par rapport aux contraintes structurelles. La dégradation de performances évaluée sur la mesure MAep est de de -20.09% à -51.56% pour la stratégie *VSCAS_SEQ* et respectivement *SVCAS_SEQ* ; et de -29,14% à -50,67% pour la stratégie *VSCAS_SAMEPLUS* et respectivement *SVCAS_SAMEPLUS.* Ces résultats peuvent être expliqués par le fait que les utilisateurs, même experts, sont peu fiables pour la formulation des requêtes impliquant la structure des documents (Trotman *et al.*, 2006b).

L'interprétation vague des contraintes structurelles pour les éléments support et cibles (la stratégie VV) augmente la pertinence des réponses par rapport à la stratégie orientée contenu (CO). L'amélioration la plus importante est obtenue pour les premières réponses retournées sur la mesure *nxCG@10* (gain de 18,7% pour *VVCAS_SAMEPLUS* et 15,44% pour *VVCAS_SEQ*). Le gain global obtenu sur la mesure *MAep* est de 6,29% pour *VVCAS_SAMEPLUS* et de 8,71% pour *VVCAS_SEQ*.

Ce comportement est confirmé par les résultats expérimentaux obtenus dans le cadre de la tâche COS.Focused d'INEX 2005 et présentés en Annexe.

**5. Conclusion**

Nous avons présenté et évalué les performances et la pertinence d'algorithmes de recherche approximative pour la recherche des informations enfouies dans des bases de documents XML. Ces algorithmes reposent sur des mécanismes d'alignement de chemins issus d'une interprétation de la requête d'une part et des sous-structures des arbres associés aux documents indexés d'autre part.

La similarité entre une requête et la structure d'un document est évaluée grâce à l'utilisation d'une distance de Levenshtein modifiée qui intègre des heuristiques et pondérations spécifiques permettent de s'affranchir d'une connaissance complète ou rigoureuse de la structure des documents exploités. Un langage de requête simple basé sur un ensemble d'opérateurs ensemblistes ou booléens, d'opérateurs qui intègrent des heuristiques type IDF a été développé dans le cadre d'une implémention qui permet de fusionner les informations sur la structure et le contenu textuel de documents XML.

Les résultats obtenus qui s'expriment en terme de vitesse d'indexation et temps moyen de réponse laisse supposer une adéquation de l'approche proposée pour le traitement de volumes conséquents de données.

En ce qui concerne la pertinence de l'approche pour la recherche d'informations dans des bases documentaires XML, des résultats expérimentaux comparées avec les résultats officiels de la campagne d'évaluation INEX 2005 ont été présentés et analysés. L'interprétation vague des informations sur la structure des documents, et conjointement la distance d'édition proposée sur les chemins XML, augmente la pertinence des réponses retournées par le système. Les résultats sont encourageants, particulièrement pour les premières réponses retournées. Ce fait est important car les premières réponses ont une grande probabilité d'être consultées par les utilisateurs.

Les perspectives envisagées à ce jour concernent l'élargissement des fonctionnalités algorithmiques pour traiter d'autres types de données non structurées telles que les séries temporelles ou les données séquentielles. Elles concernent également la parallélisation des algorithmes d'indexation et de recherche pour l'exploitation des principes proposés dans le contexte de la gestion des grandes masses de données. Des travaux en cours visent également l'adaptation et l'implémentation des algorithmes de recherche sur une plateforme physique dédiée (ACIMD – ReMIX, http://www.irisa.fr/remix).

Remerciements

## 6. Annexe : La tâche COS.Focused d'INEX 2005 (VV, VS, SV, SS vs. CO)

La tâche COS.Focused évalue la capacité du système à trouver le meilleur élément de réponse dans un chemin donné (i.e. le chevauchement des éléments de réponse est donc interdit). Pour traiter cette contrainte, un processus en deux étapes a été mis en place :

– La pertinence d'un élément de réponse est propagée récursivement à tous ses ancêtres dans la liste de réponses en utilisant une fonction *Max*. Ceci permet de prendre en compte la pertinence des éléments de réponse issus des noeuds descendants dans le calcul de score de pertinence pour l'élément de réponse courant.

– Les meilleurs N éléments de réponses non imbriqués situés au plus haut niveau dans l'arbre XML sont sélectionnés pour être retournés dans la liste des réponses finale (i.e. *HA = Highest Ancestor*). Les autres éléments de réponse sont écartés de la liste.

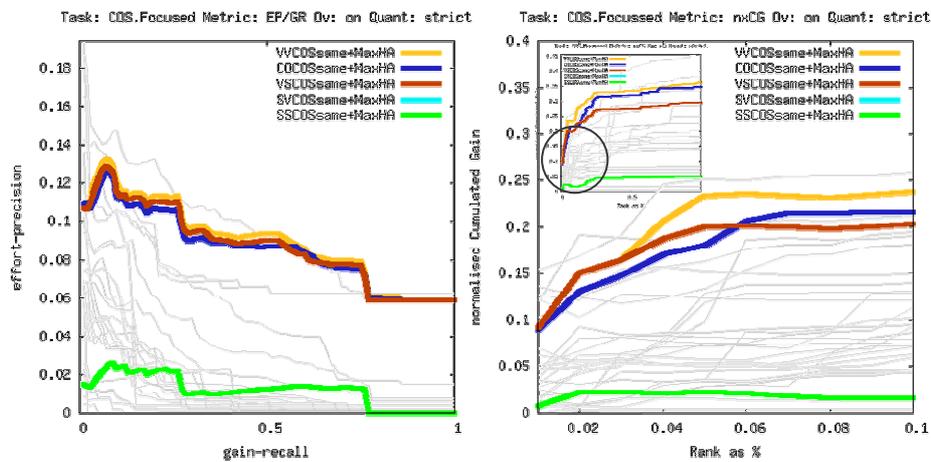

**Figure 8.** *Les courbes de rappel-précision ep/gr (gauche) et de gain cumulé étendu normalisé nxCG – zoom sur les premiers 150 réponses retournées – (droite) pour les différentes stratégies de recherche comparées avec les résultats officiels d'INEX 2005, tâche COS.Focused. Les réponses imbriquées ne sont pas autorisées (Ov.=on). Fonction d'agrégation stricte (Quant.=strict).*

| | nxCG | | | ep/gr | | | | |
|---|---|---|---|---|---|---|---|---|
| *RunId* | *@10* | *@25* | *@50* | *0,01/@15* | *0,02/@30* | *0,03/@45* | *0,1/@150* | *MAep* |
| VVCOSsame+MaxHA | **0.1059 (4)** | **0.162 (3)** | **0.2021 (2)** | **0.1086** | 0.1081 | **0.1121** | **0.1197** | **0.0891 (2)** |
| VSCOSsame+MaxHA | 0.1059 | 0.162 | 0.1853 | 0.1067 | 0.1062 | 0.1099 | 0.1154 | 0.0865 |
| COCOSsame+MaxHA | 0.1 | 0.1321 | 0.1668 | 0.109 | **0.1092** | 0.1071 | 0.1132 | 0.0852 |
| SVCOSsame+MaxHA | 0.0118 | 0.0235 | 0.0224 | 0.015 | 0.0139 | 0.0134 | 0.023 | 0.0115 |
| SSCOSsame+MaxHA | 0.0118 | 0.0235 | 0.0224 | 0.015 | 0.0139 | 0.0134 | 0.023 | 0.0115 |

**Tableau 5.** *nxCG[i] pour un rang i donné et l'effort-précision (ep) à un niveau donné de gain-rappel (gr) pour les différentes stratégies de recherche – tâche COS.Focused. Les réponses imbriquées ne sont pas autorisées (Ov.=on). Fonction d'agrégation stricte (Quant.=strict). Les rangs sur 27 soumissions sont donnés entre parenthèses. Les meilleurs résultats sont présentés en gras.*

L'évaluation des différentes stratégies (Figure 8, Tableau 5) et les comparaisons (Tableau 6) sont effectuées sur un ensemble de 19 requêtes[12] COS (i.e. des requêtes de type CAS pour lesquelles une correspondance directe à des requêtes de type CO soumises par les participants existe). Dans ce cas, la stratégie COCOS prend en compte uniquement les informations sur le contenu des éléments.

| | | nxCG | | | ep/gr | | | | |
|---|---|---|---|---|---|---|---|---|---|
| | | *@10* | *@25* | *@50* | *0,01/@15* | *0,02/@30* | *0,03/@45* | *0,1/@150* | *MAep* |
| VV, VS, SV, SS vs CO | VV | 5.9 | 22.63 | 21.16 | -0.37 | -1.01 | 4.67 | 5.74 | 4.58 |
| | VS | 5.9 | 22.63 | 11.09 | -2.11 | -2.75 | 2.61 | 1.94 | 1.53 |
| | SV, SS | -88.2 | -82.21 | -86.57 | -86.24 | -87.27 | -87.49 | -79.68 | -86.5 |

**Tableau 6.** *Gain en pourcentage introduit par les différents degrés d'approximation utilisés pour l'interprétation des contraintes structurelles (VV, VS, SV, SS) par rapport à la stratégie orientée contenu (CO) sur la tâche COS.Focused d'INEX 2005.*

## 7. Bibliographie

---

[12] INEX 2005 Topics : 202, 203, 205, 207, 208, 210, 212, 216, 219, 222, 223, 228, 229, 230, 232, 233, 234, 236, 239.